\documentclass[prl,10pt,floatfix,amsmath,twocolumn,showpacs,superscriptaddress]{revtex4-1}
\usepackage[ansinew]{inputenc}
\usepackage{graphicx}
\usepackage{pst-all}
\usepackage{color}
\usepackage{dcolumn}
\usepackage{amsmath}
\usepackage{amsthm}
\usepackage{bm}
\usepackage{layout}
\usepackage{float}
\usepackage{txfonts}
\usepackage{amsfonts}
\usepackage{amssymb}%
\usepackage[ansinew]{inputenc}
\usepackage{graphicx}
\usepackage{amsmath}
\usepackage{amsthm}
\usepackage{bm}
\usepackage{layout}
\usepackage{float}
\usepackage{amsfonts}
\usepackage{amssymb}

\def\bp{\mathbf{p}}

\def\rA{\mathrm{A}}
\def\rB{\mathrm{B}}
\def\ket#1{|#1\rangle}
\def\bra#1{\langle#1|}

\usepackage{bm}
\usepackage{graphicx}

\begin{document}

\title{Necessary and sufficient condition for non-zero quantum discord}

\author{Borivoje Daki\'c}
\affiliation{Faculty of Physics, University of Vienna,
Boltzmanngasse 5, A-1090 Vienna, Austria}
\author{Vlatko Vedral}
\affiliation{Centre for Quantum Technologies, National University of
Singapore, Singapore} \affiliation{Department of Physics, National
University of Singapore, Singapore} \affiliation{Clarendon
Laboratory, University of Oxford, Oxford UK}
\author{{\v C}aslav Brukner}
\affiliation{Faculty of Physics, University of Vienna,
Boltzmanngasse 5, A-1090 Vienna, Austria} \affiliation{Institute of
Quantum Optics and Quantum Information, Austrian Academy of
Sciences, Boltzmanngasse 3, A-1090 Vienna, Austria}

\date{\today}

\begin{abstract}

Quantum discord characterizes ``non-classicality'' of correlations
in quantum mechanics. It has been proposed as the key resource
present in certain quantum communication tasks and quantum
computational models without containing much entanglement. We obtain
a necessary and sufficient condition for the existence of non-zero
quantum discord for any dimensional bipartite states. This condition
is easily experimentally implementable. Based on this, we propose a
geometrical way of quantifying quantum discord. For two qubits this
results in a closed form of expression for discord. We apply our
results to the model of deterministic quantum computation with one
qubit (DQC1), showing that quantum discord is unlikely to be the reason
behind its speedup.

\end{abstract}


\maketitle

\emph{Introduction.}---
Quantum states of a composite system can be
divided into entangled and separable once. Entangled states display
``nonlocal features'' violating Bell's inequalities~\cite{Bell} and
are considered a necessary resource for quantum communication and pure
quantum computation allowing computational speedup over the best classical algorithm~\cite{Nielsen&Chuang}.
On the contrary, separable states are generally considered as purely
classical, since they do not violate Bell's inequalities and can be
prepared by local operations and classical communication. However,
it is valid to ask if highly mixed states, and in particular
separable states, are completely useless from quantum information
perspective. Recent investigations give compelling evidences that
this is not the case. A highly mixed state in the DQC1 model~\cite{Knill} is believed
to perform a task exponentially faster than any classical algorithm
(``without containing much entanglement''). Furthermore, it has been
shown that even some separable states contain nonclassical
correlations~\cite{Zurek,Vedral} and can create an advantage for
computing and information processing tasks over their classical
counterparts~\cite{Schack,Meyer,Caves1,Vidal,Caves2,White}.

The ``non-classicality'' of bipartite correlations is measured via
\emph{quantum discord}~\cite{Zurek}-- the discrepancy between
quantum versions of two classically equivalent expressions for
mutual information. Recently, it has been shown that almost all
quantum states have non-vanishing discord~\cite{Acin}. Quantum
discord was proposed as a figure of merit for characterizing the
nonclassical resources present in the DQC1~\cite{Caves2}. It has
been shown that initial zero-discord system-environment state is
necessary and sufficient condition for completely-positive map
evolution of the system when the environment is traced
out~\cite{Lidar,Rosario}. Furthermore, in Ref. ~\cite{Piani} is
demonstrated that if the state can be locally broadcasted than it
has vanishing discord.

Despite increasing evidences for relevance of quantum discord in
describing non-classical resources in information processing, there
is no straightforward criterion to verify the presence of discord in
a given quantum state. Its evaluation involves optimization
procedure and analytical results are known only in a few
cases~\cite{disc_evol}. In
this Letter we derive the necessary and sufficient condition for
non-vanishing quantum discord. The criterion is simple and also
experimentally friendly, since it can be evaluated directly from a
(sub)set of measurements standardly used for quantum state
tomography. Based on this, we introduce the geometrical measure of
discord and derive an explicit expression for the case of two
qubits. Finally, we give arguments putting in question
appropriateness of quantum discord to describe the non-classical
resource in DQC1 computational model.

\emph{Quantum discord.}--- Correlations between two random variables
of classical systems $A$ and $B$ are in information theory
quantified by the mutual information $I(A:B)=H(A)+H(B)-H(A,B)$. If
$A$ and $B$ are classical systems, than $H(.)$ stands for the
Shannon entropy $H(\bp)=-\sum_ip_i\log p_i$, where
$\bp=(p_1,p_2,\dots)$ is the probability distribution vector, while
$H(.,.)$ is the Shannon entropy of the joint probability
distribution $p_{ij}$. For quantum systems $A$ and $B$, function
$H(.)$ denotes the von Neumann entropy
$H(\rho)=-\mathrm{Tr}\rho\log\rho$ where $\rho$ is the density
matrix. In the classical case, we can use the Bayes rule and find an
equivalent expression for the mutual information
$I(A:B)=H(A)-H(A|B)$ where $H(A|B)$ is the Shannon entropy of $A$
conditioned on the measurement outcome on $B$. For quantum systems,
this quantity is different from the first expression for the mutual
information and the difference defines the quantum discord.

Consider a quantum composite system defined by the Hilbert space
$\mathcal{H}_{\mathrm{AB}}=\mathcal{H}_{\rA}\otimes\mathcal{H}_{\mathrm{B}}$.
Let dimensions of the local Hilbert spaces be
$\mathrm{dim}\mathcal{H_{_\rA}}=d_{\rA}$ and
$\mathrm{dim}\mathcal{H_{_\mathrm{B}}}=d_{\mathrm{B}}$, while
$d=\mathrm{dim}\mathcal{H_{_\mathrm{AB}}}=d_{\rA}d_{\mathrm{B}}$ .
Given a state $\rho$ (density matrix) of a composite system,
the total amount of correlations is quantified by quantum mutual
information~\cite{Winter}:
\begin{equation}
I(\rho)=H(\rho_{\rA})+H(\rho_{\rB})-H(\rho),
\end{equation}
where $H(\rho)$ is the von Neumann entropy and
$\rho_{\rA,\mathrm{B}}=\mathrm{Tr}_{\rB,\mathrm{A}}(\rho)$ are
reduced density matrices. A generalization of the classical
conditional entropy is $H(\rho_{\mathrm{B}|\rA})$, where
$\rho_{\mathrm{B}|\rA}$ is the state of $B$ given a measurement on
$A$. By optimizing over all possible measurements in $A$, we define
an alternative version of the mutual information
\begin{equation}
Q_{A}(\rho)=H(\rho_{\mathrm{B}})-\mathrm{min}_{\{E_k\}}\sum_k p_k
H(\rho_{\mathrm{B}|k}),
\end{equation}
where
$\rho_{\mathrm{B}|k}=\mathrm{Tr}_{\rA}(E_k\otimes\openone_{\mathrm{B}}\rho)/\mathrm{Tr}(E_k\otimes\openone_{\mathrm{B}}\rho)$
is the state of $B$ conditioned on outcome $k$ in $A$ and $\{E_k\}$
represents the set of positive operator valued measure elements. The discrepancy between the two
measures of information defines the quantum discord ~\cite{Zurek,Vedral}:
\begin{equation}\label{dicord}
D_{A}(\rho)=I(\rho)-Q_{A}(\rho).
\end{equation}
The discord is always nonnegative~\cite{Zurek} and reaches zero for
the classically correlated states~\cite{Vedral}. Note that discord
is not a symmetric quantity $D_{A}(\rho)\neq D_{B}(\rho)$ and $D_A$
refers to the ``left'' discord, while $D_B$ refers to the ``right''
discord. The state $\rho$ for which $D_{A}(\rho)=D_{B}(\rho)=0$ is
\emph{completely} classically correlated in a sense
of~\cite{Oppenheim,Williamson} 
From now on, when we refer to the discord we mean the ``left''
discord $D_A$.

To give an example of a state with non-vanishing discord consider
the two-qubit separable state in which four nonorthogonal states of
one qubit are correlated with four nonorthogonal states of the
second qubit:
\begin{eqnarray}\nonumber
&&\frac{1}{4}(\ket{0}\bra{0}\otimes\ket{+}\bra{+}+\ket{1}\bra{1}\otimes\ket{-}\bra{-}+\ket{+}\bra{+}\otimes\ket{1}\bra{1}\\
&&+\ket{-}\bra{-}\otimes\ket{0}\bra{0}).
\end{eqnarray}

Unlike the state above, one can show that the state $\rho$ is of
zero-discord if and only if there exist a von Neumann measurement
$\{\Pi_k=\ket{\psi_k}\bra{\psi_k}\}$ 
such that~\cite{Datta}
\begin{equation}\label{0discord}
\sum_k(\Pi_k\otimes\openone_\rB)\rho(\Pi_k\otimes\openone_\rB)=\rho,
\end{equation}
In other words the zero-discord state is of the form $\rho=\sum_k
p_k\ket{\psi_k}\bra{\psi_k}\otimes\rho_k$ where $\{\ket{\psi_k}\}$
is some orthonormal basis set, $\rho_k$ are the quantum states in
$B$ and $p_k$ are non-negative numbers such that $\sum_k p_k=1$.

\emph{An easily implementable necessary and sufficient condition.}--- Let us choose basis
sets in local Hilbert-Schmidt spaces of Hermitian operators,
$\{A_n\}$ and $\{B_m\}$ where $n=1\dots d_\rA^2$ and $m=1\dots
d_\rB^2$. We decompose the state $\rho$ of composite system into
$\rho=\sum_{nm}r_{nm}A_n\otimes B_m$. The coefficients $r_{nm}$ define
$d_\rA^2\times d_\rB^2$ real matrix $R$ which we call the correlation
matrix. We can find its singular value decomposition (SVD), $U R
W^{\mathrm{T}}=\mathrm{diag}[c_1,c_2,\dots]$ where $U$ and $W$ are
$d_\rA^2\times d_\rA^2$ and $d_\rB^2\times d_\rB^2$ orthogonal
matrices, respectively, while $\mathrm{diag}[c_1,c_2,\dots]$ is
$d_\rA^2\times d_\rB^2$ diagonal matrix. SVD defines new basis in
local Hilbert-Schmidt spaces $S_{n}=\sum_{n'}U_{nn'}A_{n'}$ and
$F_{m}=\sum_{m'}W_{mm'}B_{m'}$. The state $\rho$ in the new basis is of
the form $\rho=\sum_{n=1}^{L}c_n S_n\otimes F_n$ where
$L=\mathrm{rank}R$ is the rank of correlation matrix $R$ (the number
of non-zero eigenvalues $c_n$).

The necessary and sufficient condition (\ref{0discord}) becomes
$\sum_{n=1}^{L}c_n (\sum_k\Pi_kS_n\Pi_k)\otimes
F_n=\sum_{n=1}^{L}c_n S_n\otimes F_n$ and it is equivalent to the
set of conditions:
\begin{equation}
\sum_k\Pi_k S_n\Pi_k=S_n,~~~n=1\dots L,
\end{equation}
or equivalently $[S_n,\Pi_k]=0$ for all $k,n$. This means that the
set of operators $\{S_n\}$ have common eigenbasis defined by the set
of projectors $\{\Pi_k\}$. Therefore, the set $\{\Pi_k\}$ exists if
and only if:
\begin{equation}\label{NSC}
[S_n,S_m]=0,~~n,m=1\dots L.
\end{equation}
In order to show zero discord we have to check at most $L(L-1)/2$
commutators, where
$L=\mathrm{rank}R\leq\mathrm{min}\{d_\rA^2,d_\rB^2\}$. Now, recall
that the state of zero discord is of the form
$\rho=\sum_{k=1}^{d_\rA} p_k\Pi_k\otimes\rho_k$, therefore is a sum
of at most $d_\rA$ product operators. This bounds the rank of the
correlation tensor to $L\leq d_\rA$. Thus, the rank of the
correlation tensor is the simple discord witness: If $L>d_\rA$, the
state has a non-zero discord.

Correlation matrix can be obtained directly by simple measurements
usually involved in quantum state tomography. However, the detection
of non-zero discord does not necessarily require measurement of all
$(d_Ad_B)^2$ elements of the correlation matrix (full state
tomography). It is sufficient that the experimentalist measures that
many elements of the correlation matrix until he finds $d_\rA+1$
linearly independent rows (or columns) of the correlation matrix.

\emph{Geometric measure of discord.}--- Evaluation of quantum
discord given by equation (\ref{dicord}) in general requires
considerable numerical minimization. Different measures of quantum
discord~\cite{Terno} and their extensions to multipartite
systems~\cite{Williamson} have been proposed. However, analytical
expression are known only for certain classes of
states~\cite{disc_evol}. Here
we propose a following geometric measure
\begin{equation}\label{GMD}
D_A^{(2)}(\rho)=\mathrm{min}_{\chi\in\Omega_0}||\rho-\chi||^2,
\end{equation}
where $\Omega_0$ denotes the set of zero-discord states and
$||X-Y||^2=\mathrm{Tr}(X-Y)^2$ is the square norm in the
Hilbert-Schmidt space. We will show how to evaluate this quantity
for an arbitrary two-qubit state.

\emph{Two-qubit case.}--- Consider the case
$\mathcal{H}_{\rA}=\mathcal{H}_{\mathrm{B}}=\mathbb{C}^2$. We write
a state $\rho$ in Bloch representation:
\begin{equation}
\rho=\frac{1}{4}(\openone\otimes\openone+\sum_{i=1}^3x_i\sigma_i\otimes\openone+\sum_{i=1}^3y_i\openone\otimes\sigma_i+
\sum_{i,j=1}^3T_{ij}\sigma_i\otimes\sigma_j),
\end{equation}
where $x_i=\mathrm{Tr}\rho(\sigma_i\otimes\openone)$,
$y_i=\mathrm{Tr}\rho(\openone\otimes\sigma_i)$ are components of the
local Bloch vectors,
$T_{ij}=\mathrm{Tr}\rho(\sigma_i\otimes\sigma_j)$ are components of
the correlation tensor, and $\sigma_i$,  $i \in \{1,2,3\}$, are the
three Pauli matrices. To each  state $\rho$ we associate the triple
$\{\vec{x},\vec{y},T\}$. Now, we characterize the set $\Omega_0$. A
zero-discord state is of the form
$\chi=p_1\ket{\psi_1}\bra{\psi_1}\otimes\rho_1+p_2\ket{\psi_2}\bra{\psi_2}\otimes\rho_2$,
where $\{\ket{\psi_1},\ket{\psi_2}\}$ is a single-qubit orthonormal
basis, $\rho_{1,2}$ are $2\times2$ density matrices, and $p_{1,2}$
are non-negative numbers such that $p_1+p_2=1$. We define
$t=p_1-p_2$ and three vectors
\begin{eqnarray}
\vec{e}&=&\bra{\psi_1}\vec{\sigma}\ket{\psi_1},\\
\vec{s}_{\pm}&=&\mathrm{Tr}(p_1\rho_1\pm
p_2\rho_2)\vec{\sigma}.
\end{eqnarray}
It can easily be shown that $t\vec{e}$ and $\vec{s}_{+}$ represent
the local Bloch vectors of the first and second qubit, respectively,
while the vector $\vec{s}_{-}$ is directly related to the
correlation tensor which is of the product form
$T=\vec{e}\vec{s}_{-}^{\mathrm{T}}$. Therefore, a state of
zero-discord $\chi$ has Bloch representation
$\vec{\chi}=\{t\vec{e},\vec{s}_{+},\vec{e}\vec{s}_{-}^{\mathrm{T}}\}$,
where $||\vec{e}||=1$, $||\vec{s}_{\pm}||\leq1$ and $t\in[-1,1]$.
The distance between states $\rho$ and $\chi$ is given by
\begin{eqnarray}\label{func}
||\rho-\chi||^2&=&||\rho||^2-2\mathrm{Tr}\rho\chi+||\chi||^2\\\nonumber
&=&\frac{1}{4}(1+||\vec{x}||^2+||\vec{y}||^2+||T||^2)\\\nonumber
&-&\frac{1}{2}(1+t\vec{x}\vec{e}+\vec{y}\vec{s}_{+}+\vec{e}T\vec{s}_{-})\\\nonumber
&+&\frac{1}{4}(1+t^2+||\vec{s}_{+}||^2+||\vec{s}_{-}||^2),
\end{eqnarray}
where $||T||^2=\mathrm{Tr}T^{\mathrm{T}}T$. First, we optimize the
distance over parameters $\vec{s}_{\pm}$ and $t$. The function of equation (\ref{func}) is convex and quadratic in its variables $t, \vec{s}_{\pm}$. It is straightforward to see that its Hessian is a positive and non-singular matrix. Therefore the function has a unique global minimum. The minimum occurs
when the derivative is zero:
\begin{eqnarray}
\frac{||\rho-\chi||^2}{\partial
t}&=&\frac{1}{2}(-\vec{x}\vec{e}+t)=0,\\
\frac{||\rho-\chi||^2}{\partial
\vec{s}_{+}}&=&\frac{1}{2}(-\vec{y}+\vec{s}_{+})=0,\\
\frac{||\rho-\chi||^2}{\partial
\vec{s}_{-}}&=&\frac{1}{2}(-T^{\mathrm{T}}\vec{e}+\vec{s}_{-})=0,
\end{eqnarray}
which gives the solution $t=\vec{x}\vec{e}$, $\vec{s}_{+}=\vec{y}$
and $\vec{s}_{-}=T^{\mathrm{T}}\vec{e}$. Since the solution lies within the range of parameters, $|\vec{x}\vec{e}|,||\vec{y}||,|| T^{\mathrm{T}}\vec{e}||\leq1$ it represents the global minimum. After substituting the
solution we obtain
$||\rho-\chi||^2=\frac{1}{4}\left(||\vec{x}||^2+||T||^2-\vec{e}(\vec{x}\vec{x}^{\mathrm{T}}+TT^{\mathrm{T}})\vec{e}\right)$
which attains the minimum when $\vec{e}$ is an eigenvector of matrix
$K=\vec{x}\vec{x}^{\mathrm{T}}+TT^{\mathrm{T}}$ for the largest
eigenvalue. Therefore, we have:
\begin{equation}
D_A^{(2)}(\rho)=\frac{1}{4}(||\vec{x}||^2+||T||^2-k_{\mathrm{max}}),
\end{equation}
where $k_{\mathrm{max}}$ is the largest eigenvalue of matrix
$K=\vec{x}\vec{x}^{\mathrm{T}}+TT^{\mathrm{T}}$. Next, we apply our
criterion to a class of states.

\begin{figure}\centerline{
\includegraphics[width=7cm]{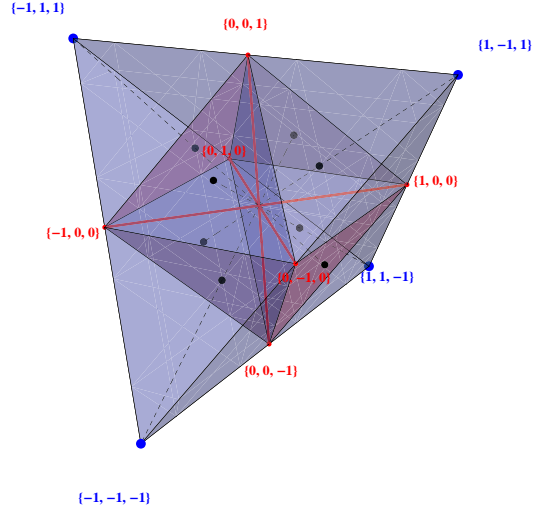}}
\caption{ The set of two-qubit states with maximally mixed marginals
(i.e. the reduced states of individual qubits are completely mixed).
Physical states belong to the tetrahedron, among which separable
ones are confined to the octahedron. The zero-discord states are
labeled by the red lines (it is therefore clear that almost all
states have non-zero discord~\cite{Acin}). The states with maximal
value of discord correspond to the vertices of the tetrahedron (the
four Bell states). Among the set of separable states, those which
maximize discord are the centers of octahedron facets
$(\pm1,\pm1,\pm1)/3$ (black dots).}\label{discordPlot}
\end{figure}

\emph{States with maximally mixed marginals.}--- We consider an
example of two qubit states with maximally mixed marginals. Such a
state is locally equivalent (under some local unitary transformation
$U_1\otimes U_2$) to a state
$\rho(\vec{t})=(\openone\otimes\openone+\sum_{i=1}^3t_{i}\sigma_i\otimes\sigma_i)/4$,
where $\vec{t}=(t_1,t_2,t_3)$. The state $\rho(\vec{t})$ is physical
if $\vec{t} $ belongs to the tetrahedron (Figure \ref{discordPlot})
defined by the set of vertices $(-1,-1,-1)$, $(-1,1,1)$, $(1,-1,1)$
and $(1,1,-1)$, while is separable if $\vec{t}$ belongs to the
octahedron defined by the set of vertices $(\pm1,0,0)$, $(0,\pm1,0)$
and $(0,0,\pm1)$~\cite{Horodecki}. Simple calculation shows that
$D_A^{(2)}(\vec{t})=\frac{1}{4}(t_1^2+t_2^2+t_3^2-\mathrm{max}\{t_1^2,t_2^2,t_3^2\})$.
The zero-discord states have at most one non-zero component of
vector $\vec{t}$ (Figure \ref{discordPlot}, red lines). The function
$D_A^{(2)}(\vec{t})$ reaches its maximal value of $D_A^{(2)}=1/2$ at
the vertices of tetrahedron which represent the four Bell states. Within the set of separable states
(octahedron) its maximal value of $D_A^{(2)}=1/6$ is attained at the
centers of octahedron facets $(\pm1,\pm1,\pm1)/3$. They represent
the states
\begin{equation}
\rho_{i_1i_2i_3}=\frac{1}{4}(\openone\otimes\openone+\frac{1}{3}\sum_{k=1}^3(-1)^{i_k}\sigma_k\otimes\sigma_k),
\end{equation}
where $i_k=\pm1$, and can intuitively be understood as equal mixture
of ``maximally non-orthogonal'' states. The states are symmetric
under exchange of subsystems, thus they have the same value of
``left'' and ``right'' discord $D_A=D_B$.

\begin{figure}\centerline{
\includegraphics[width=6cm]{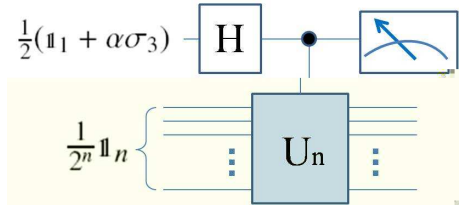}}
\caption{The quantum circuit for estimating the normalized trace of
the unitary matrix $U_n$ using the model of deterministic computing
with one quantum bit (DQC1). $H$ stands for the Hadamard gate. The
control (top) qubit is measured in the $\sigma_1$ and $\sigma_2$
basis, and the expectation values give the real and imaginary part
of normalized trace $\tau=\mathrm{Tr}U_n/2^n$ with the overhead
scaling as $\frac{1}{\alpha^2}$~\cite{Caves2}.}\label{DQC1}
\end{figure}

\emph{DQC1 model.}--- In ~\cite{Knill}, Knill and Laflamme
introduced the model of mixed-state quantum computing which preforms
the task of evaluating the normalized trace of a unitary matrix
efficiently. The corresponding quantum circuit is shown in Figure
\ref{DQC1}. The input state is a highly mixed separable state and
consists of a control qubit in the state
$\frac{1}{2}(\openone+\alpha\sigma_3)$, where $\alpha$ describes the
purity, and a collection of $n$ qubits in the maximally mixed state
$\frac{1}{2^n}\openone_n$, where $\openone_n$ is the $n$-qubit
identity. The DQC1 circuit consists of the Hadamard gate applied to
the control qubit and a control $n$-qubit unitary gate $U_n$. The output state is:
\begin{equation}\label{DQC1 state}
\rho=\frac{1}{2^{n+1}}(\openone_1\otimes\openone_n+\alpha\ket{1}\bra{0}\otimes
U_n+\alpha\ket{0}\bra{1}\otimes U_n^{\dagger}).
\end{equation}
We consider only the cases $\alpha\neq0$, otherwise the state at the
output is completely mixed and therefore cannot accomplish the task.
After measuring the control qubit at the output in the eigenbasis of
$\sigma_1$ and $\sigma_2$, we retrieve the normalized trace of the
unitary matrix $\tau=\mathrm{Tr}U_n/2^n$ with the polynomial
overhead scaling $1/\alpha^2$~\cite{Caves2}.

The control qubit is completely separable from the rest of the
qubits. The output state has vanishingly small entanglement across
any bipartite split that groups the control qubit with some of the
mixed qubits~\cite{Knill}. However, there is strong evidence that
DQC1 task cannot be preformed efficiently using classical
computation~\cite{Vidal}. The question is what brings a ``speed-up''
in the considered task? The quantum discord was proposed as a figure
of merit for characterizing the resources present in DQC1
model~\cite{Caves2}. It has been shown that for almost every unitary
matrix $U_n$ (random unitary) the discord in the output state
(\ref{DQC1 state}) is non-vanishing. Here we derive an explicit
condition for characterizing the correlations in the output state
and show that the discord is unlikely to be the source of speedup.
We re-write it into a form:
\begin{equation}\label{DQC1 state1}
\rho=\frac{1}{2^{n+1}}\left(\openone_1\otimes\openone_n+\alpha\sigma_1\otimes
\frac{U_n+U_n^{\dagger}}{2}+\alpha\sigma_2\otimes\frac{U_n-U_n^{\dagger}}{2i}\right).
\end{equation}
Now, we apply the condition (\ref{NSC}). The operators $\sigma_1$
and $\sigma_2$ do not commute, therefore, the state $\rho$ is of the
zero-discord if and only if the operators
$\frac{U_n+U_n^{\dagger}}{2}$ and $\frac{U_n-U_n^{\dagger}}{2i}$ are
linearly dependent, or equivalently $U_n^{\dagger}=k U_n$. This is
possible if and only if $U_n=\mathrm{e}^{i\phi}A$, where $A^2=\openone$ is
a binary observable. For such a unitary all the correlations at the
output of DQC1 circuit are classical. However, it is very unlikely
that the normalized trace of $\mathrm{e}^{i\phi}A$ can be evaluated
efficiently on a classical computer, since all it's eigenvectors can
be arbitrarily complex (random states).

We emphasize that our measure of discord is not monotonic under
local operations. This, however, is not a shortcoming, as discord,
unlike entanglement and mutual information, can in fact increase as
well as decrease under local operations (even without the presence
of classical correlations). A simple example of the local increase
is to start from a zero-discord state $|00\rangle\langle
00|+|11\rangle\langle 11|$ and transform, say the first qubit, so
that $|0\rangle \rightarrow |\psi_0\rangle$ and $|1\rangle
\rightarrow |\psi_1\rangle$, such that $|\psi_0\rangle$ and
$|\psi_1\rangle$ are not orthogonal. The resulting state,
$|0\Psi_0\rangle\langle 0\Psi_0|+|1\psi_1\rangle\langle 1\psi_1|$
clearly has a non-vanishing discord. Finally, we point out that our
method can be extended to any number of subsystems, though
evaluating the measure of discord becomes progressively more
difficult with increasing number of subsystems and their
dimensionality.

\acknowledgements

We acknowledge support from the Austrian Science
Foundation FWF within Project No. P19570-N16, SFB
and CoQuS No. W1210-N16, and the European
Commission Project QAP (No. 015848). V.V. acknowledges
financial support from the National Research
Foundation and Ministry of Education in Singapore and
the support of Wolfson College Oxford.

\emph{Note added in proof.}---A related work was done by Datta~\cite{Datta1}.

\end{document}